\documentclass[aps,prl,showpacs,reprint,english,twocolumn,amsmath,amssymb,superscriptaddress,nobibnotes]{revtex4-1}

\usepackage{amsmath}
\usepackage{graphicx}
\usepackage{amssymb}
\begin{document}

%\preprint{}

\title{Experimental evidence of quantum quenching of fluctuations in heavy ion collisions}



\author{B.C. Stein}
\email[Corresponding Author. Email:  ]{stein@comp.tamu.edu}
\affiliation{Cyclotron Institute, Texas A\&M University, College Station, TX 77843, USA}
\affiliation{Department of Chemistry, Texas A\&M University, College Station, TX 77843, USA}

\author{H. Zheng}
\affiliation{Cyclotron Institute, Texas A\&M University, College Station, TX 77843, USA}
\affiliation{Department of Physics, Texas A\&M University, College Station, TX 77843, USA}

\author{A. Bonasera}
\affiliation{Cyclotron Institute, Texas A\&M University, College Station, TX 77843, USA}
\affiliation{Laboratori Nazionali del Sud, INFN, via Santa Sofia, 62, 95123 Catania, Italy}

\author{S.N. Soisson}
\altaffiliation[Present Address: ]{Sandia National Laboratories, Albuquerque, NM 87123, USA}
\affiliation{Cyclotron Institute, Texas A\&M University, College Station, TX 77843, USA}
\affiliation{Department of Chemistry, Texas A\&M University, College Station, TX 77843, USA}

\author{A. McIntosh}
\affiliation{Cyclotron Institute, Texas A\&M University, College Station, TX 77843, USA}

\author{R. Tripathi}
\altaffiliation[Present Address:  ]{Radiochemistry Division, Bhabha Atomic Research Center, Mumbai, India}
\affiliation{Cyclotron Institute, Texas A\&M University, College Station, TX 77843, USA}

\author{G.A. Souliotis}
\affiliation{Cyclotron Institute, Texas A\&M University, College Station, TX 77843, USA}
\affiliation{Laboratory of Physical Chemistry, Department of Chemistry, National and Kapodistrian University of Athens, Athens 15771, Greece}

\author{P. Marini}
\altaffiliation[Present Address:  ]{Grand AccŽlŽrateur National d'Ions Lourds, Bd Henri Becquerel, BP 55027 - 14076 CAEN Cedex 05, France}
\affiliation{Cyclotron Institute, Texas A\&M University, College Station, TX 77843, USA}

\author{A.L. Keksis}
\altaffiliation[Present Address:  ]{C-NR, Los Alamos National Laboratory, Los Alamos, New Mexico, 87545, USA}
\affiliation{Cyclotron Institute, Texas A\&M University, College Station, TX 77843, USA}
\affiliation{Department of Chemistry, Texas A\&M University, College Station, TX 77843, USA}

\author{S. Wuenschel}
\affiliation{Cyclotron Institute, Texas A\&M University, College Station, TX 77843, USA}
\affiliation{Department of Chemistry, Texas A\&M University, College Station, TX 77843, USA}

\author{Z. Kohley}
\altaffiliation[Present Address:  ]{National Superconducting Cyclotron Laboratory, Michigan State University, East Lansing, Michigan 48824, USA}
\affiliation{Cyclotron Institute, Texas A\&M University, College Station, TX 77843, USA}
\affiliation{Department of Chemistry, Texas A\&M University, College Station, TX 77843, USA}

\author{L.W. May}
\affiliation{Cyclotron Institute, Texas A\&M University, College Station, TX 77843, USA}
\affiliation{Department of Chemistry, Texas A\&M University, College Station, TX 77843, USA}

\author{S.J. Yennello}
\affiliation{Cyclotron Institute, Texas A\&M University, College Station, TX 77843, USA}
\affiliation{Department of Chemistry, Texas A\&M University, College Station, TX 77843, USA}


\date{\today}

\begin{abstract}
The first experimental results of a new quantum method for calculating nuclear temperature and density of fragmenting heavy ions is presented.  This method is based on fluctuations in the event quadrupole momentum and fragment multiplicity distributions of light Fermions.  The calculated temperatures are lower than those obtained with a similar classical method.  Quenching of the normalized multiplicity distributions of light fermions due to Pauli blocking is also observed.  These results indicate a need for a quantum treatment when dealing with statistical properties of fragmenting heavy-ions.  
\end{abstract}

%\keywords{}

\maketitle


%\section{}

The experimental measurement of temperature and density has significant impact on the field of intermediate energy heavy-ion reactions.  Constraining the nuclear equation of state and investigations of a liquid-gas phase transition in nuclear matter are two areas of intense interest that rely heavily on such quantities.  Of broader interest, the nucleus itself provides a laboratory to look at the thermal properties of a quantum system.  It was recognized some time ago by W. Bauer that it would be necessary to correct for the Fermi motion in probes derived from the kinetic energy of emitted fragments \cite{bauer1995}.  Despite this, most of the commonly used temperature calculations are derived from classical statistical mechanics \cite{morrissey1994,kelic2006}.  One exception is a recently proposed calculation based on fluctuations in the event quadrupole momentum and particle multiplicity distributions \cite{zheng2011,zhengArxiv}.  A similar method has been used with trapped Fermi gases, which shows a quenching in the density fluctuations of the gas due to Pauli blocking \cite{muller2010,sanner2010,westbrook2010}.  An analogous signal should be observable in the fragmentation of heavy ions.  In this letter, we will present the first experimental quantum temperatures calculated by the method presented in Ref. \cite{zheng2011}.  Qualitative comparisons will be made to predictions of the Constrained Molecular Dynamics Model (CoMD), a microscopic model that includes Fermionic statistics, for a similar fragmenting system \cite{bonasera2000,papa2001}.  The experimental effect of Pauli blocking on the particle multiplicity distributions of light fermions will also be presented.

The experimental data was acquired at the Cyclotron Institute of Texas A\&M University.  A beam of 45 MeV/A $^{32}$S ions was accelerated by the K500 superconducting cyclotron and then impinged upon a highly enriched $^{112}$Sn target.  Charged particles from the reaction were measured by the FAUST detector array.  FAUST is a segmented array of 68 Si \-- CsI(Tl) telescopes with coverage from 1.6 to 44.8 degrees with respect to the beam axis.  The array is structured into 5 square rings with the longitudinal distance from the silicon detector to the target kept constant within each ring.  This arrangement gives FAUST solid angle coverage of 90\% from 2.3 to 33.6 degrees, 71\% from 1.6 to 2.3 degrees, and 25\% from 33.6 to 44.8 degrees.  A detailed description of FAUST may be found in Ref. \cite{gimeno97,stein2009}.  It should be noted that the experimental setup was not designed to be sensitive to photons or neutrons; only charged particles were detected.  

Particle identification was performed using the $\Delta$E\--E technique.  For details see Ref. \cite{stein2009}.  Such identification requires the fragment to have sufficient energy to pass through the silicon detector before implanting in the CsI(Tl).  Because of the thickness of the silicon, the charged particles collected are, by design, heavily biased towards fragments produced from projectile-like sources (PLS) in peripheral and mid-peripheral collisions.  Particles were identified isotopcially (in both atomic charge and mass number) for particles with a charge up to at least $Z=8$ throughout the array and  $Z=13$ for select detectors.  Particles for which the mass could not be determined were identified in charge only up to the charge of the incident beam ($Z=16$).

PLS were reconstructed on an event-by-event basis.  Events were selected based on two constraints.  First, the sum of the charge of detected particles was required to be the charge of the beam (Z=16).  This condition has previously been shown to select for events with properties indicative of a projectile-like source \cite{keksis2010,soisson2011}.  Second, the sum of the mass of the detected particles was required to be the mass of the beam (A=32).  This also implies that all fragments in the event are mass identified and that the minimum charged particle multiplicity per event is two.  By fixing the source to a single charge and mass, effects due to differences in the size or isospin asymmetry of the source should be significantly reduced.  

Event excitation was calculated by calorimetry.  In order to minimize dynamical contributions to the kinetic energy  \cite{wuenschel2010,wuenschelThesis}, only the components of the momentum transverse to the PLS-TLS (target-like source) separation axis were used in the calculation:
\begin{eqnarray}
K=(3/2)\frac{(p_{x}^{2}+p_{y}^{2})}{2m}
\end{eqnarray}
where K is the kinetic energy, $p_{x}^{2}$ and $p_{y}^{2}$ are the transverse components of a fragment's momentum in the frame of the PLS, and m is the fragment mass.  The event excitation energy was calculated: 
\begin{eqnarray}
E_{xy}^{*}=\sum_{i}^{}(K_{i}+\Delta M_{i})-\Delta M_{PLS}
\end{eqnarray}
where K$_{i}$ is the kinetic energy of a fragment in the frame of the PLS, $\Delta$M$_{i}$ is the mass excess of a given fragment and $\Delta$M$_{PLS}$ is the mass excess of the PLS.  The summation is over all detected charged particles in an event.  Events were sorted into 12 bins in event excitation energy, each bin has a width of 0.5 MeV.  It should be noted that this estimate represents an apparent excitation since the experimental setup was not sensitive to emitted neutrons.      

%%%-----------------------------------------------------------------------------
%%% Fig. 1
%%%---------
\begin{figure}[tbp]
\includegraphics[width=0.50\textwidth, angle=0]{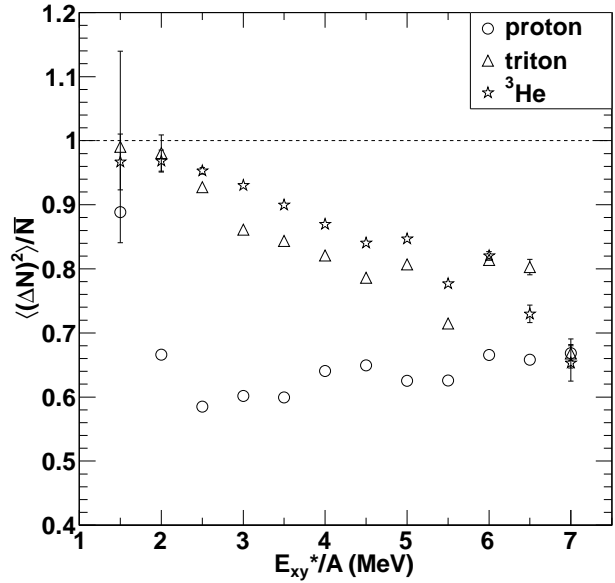}
\caption{Normalized multiplicity versus event excitation energy is plotted for protons, tritons, and $^{3}$He.  Notice that in all three cases the normalized multiplicity shows quenching of the multiplicity fluctuations compared to the classical limit ($\left<(\Delta N)^{2}\right>/\bar N=1$) denoted by a dashed line.  Statistical error is plotted for the normalized multiplicity and the excitation bins are 0.5 MeV in width.}
\label{fig:mult}
\end{figure}
%%%-------------------------------------------------------------------------------

The temperature calculation is based on the fluctuations in the event quadrupole moment.  Details of this method can be found in reference \cite{zheng2011}.  Similar to the excitation energy calculation discussed above, any dynamical contribution to the fragment kinetic energy would largely affect the longitudinal component. The quadrupole momentum is therefore defined as $Q_{xy}=<p_{x}^{2}-p_{y}^{2}>$ where $p_{x}$ and $p_{y}$ are the same transverse components of a given fragment's momentum used in the event excitation calculation.  The variance of the quadrupole is then given by:
\begin{eqnarray}
\sigma_{xy}^{2}=\int d^{3}p(p_{x}^{2}-p_{y}^{2})f(p)
\end{eqnarray}
where $f(p)$ is the momentum distribution of the particles.  In earlier work, a classical Maxwell-Boltzmann distribution was assumed for $f(p)$ giving the variance as:
\begin{eqnarray}
\sigma_{xy}^{2}=\bar{N}4m^{2}T_{cl}^{2}
\end{eqnarray}
where $T_{cl}$ is the classical temperature, m is the mass of the fragment, and $\bar N$ is the average multiplicity per event of the given fragment type \cite{wuenschel2010,wuenschelThesis,zheng2011}.  This classical result will be useful for later comparison to the quantum method.

For the quantum calculation, the authors of Ref. \cite{zheng2011} constrained themselves to A=1 fermions so we will do the same here.  If a Fermi-Dirac distribution is used in Eq. 3, and $\varepsilon_{f}=\varepsilon_{f_{0}}(\rho/\rho_{0})^{(2/3)}=36(\rho/\rho_{0})^{(2/3)}$ MeV is the Fermi energy of nuclear matter, using an expansion on $T/\varepsilon_{f}$ we obtain:
\begin{eqnarray}
\sigma_{xy}^{2}=\bar{N} \left[ \frac{16m^{2}\varepsilon_{f}^{2} }{35} \left( 1+\frac{7}{6}\pi^{2}\left( \frac{T}{\varepsilon_{f}}\right)^{2}+O \left( \frac{T}{\varepsilon_{f}} \right)^{4} \right)  \right]
\end{eqnarray}
This gives us the quadrupole fluctuations in terms of temperature and density through $\varepsilon_{f}$ \cite{zheng2011,zhengArxiv}.  The value of $\varepsilon_{f_{0}}=36$ MeV was chosen to be consistent with Ref. \cite{zheng2011}.

Following a similar derivation, the fluctuations of a fermion multiplicity distribution can be written as \cite{zheng2011}:
\begin{eqnarray}
\frac{\left<(\Delta N)^{2}\right>}{\bar N}=\frac{3}{2}\frac{T}{\varepsilon_{f}}+O \left( \frac{T}{\varepsilon_{f}} \right)^{3}
\end{eqnarray}
By substituting Eq. 6 into Eq. 5 the Fermi energy can be expressed in terms of quadrupole and multiplicity fluctuations.  It is then simple to calculate the quantum temperature from either Eq. 4 or Eq. 5 \cite{zheng2011}.  It should be noted that the above derivation assumes that $T/\varepsilon_{f} < 1$.

Fig. \ref{fig:mult} shows the measured normalized multiplicity fluctuations ($\left<(\Delta N)^{2}\right>/\bar N$) for protons, tritons, and $^{3}$He particles as a function of excitation energy per nucleon.  For a classical gas, the multiplicity of a specific cluster size will follow Poissonian statistics for which $\left<(\Delta N)^{2}\right>/\bar N=1$ \cite{sanner2010}.  If, instead, we have a Fermi gas, the Pauli exclusion principal prevents multiple fermions of the same type from occupying the same state.  This has the effect of blocking certain fragment emission channels leading to a reduction or quenching of the multiplicity fluctuations.  Such quenching is observed in Fig. \ref{fig:mult} for all three particle types and agrees qualitatively with the predictions of the CoMD model \cite{zhengArxiv} as well as the results from experiments with trapped Fermi gasses \cite{muller2010,sanner2010}.  The increased quenching of the protons over that of the tritons, and $^{3}$He particles is believed to be due to the composite nature of the A=3 particles.  Moreover, A=3 particles have a smaller mean multiplicity compared to that of protons, and therefore A=3 particles have a smaller probability for interacting with particles of their own type.

%%%-----------------------------------------------------------------------------
%%% Fig. 2
%%%---------
\begin{figure}[tbp]
\includegraphics[width=0.50\textwidth, angle=0]{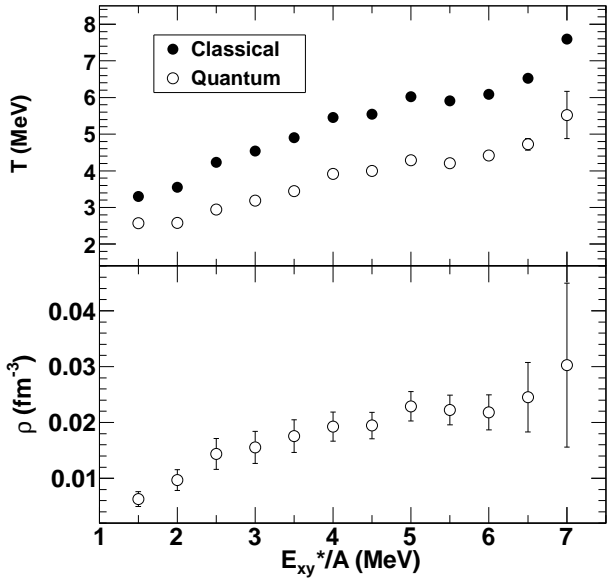}
\caption{(Top) The proton temperature is plotted as a function of event excitation energy for both the classical and quantum calculations.  The quantum method shows a significant reduction in temperature over that of the classical limit for all excitation bins.  (Bottom) The proton density as calculated through the quantum method is plotted versus event excitation energy.  In both plots, statistical error is shown and the event excitation bins match that of Fig. \ref{fig:mult}.}
\label{fig:temp}
\end{figure}
%%%-------------------------------------------------------------------------------

Since the quantum derivation is specific to A=1 fermions, only results based on the proton fluctuations will be presented from this point forward.  A comparison of the classical vs quantum temperature is plotted in the top panel Fig. \ref{fig:temp} as a function of excitation energy per nucleon.  There is a significant increase in temperature in the classical over the quantum calculation with the difference exceeding 2 MeV in the highest excitation bins.  The increased value of the classical temperature was predicted in references \cite{zheng2011,bauer1995} and is caused by classical calculations failing to account for the portion of a fragment's kinetic energy due to the Fermi motion of the nucleons prior to emission.

The bottom panel in Fig. \ref{fig:temp} shows the proton density vs event excitation energy.  The density was calculated assuming a value of $\rho_{0}=0.15$ fm$^{-3}$:  
\begin{eqnarray}
\varepsilon_{f}=36(\rho/\rho_{0})^{(2/3)} MeV
\end{eqnarray}
using the fermi energy calculated from Eq. 5 and Eq. 6.  The range of densities, from approximately 0.01 to 0.03 fm$^{-3}$ agree well with recent calculations by Wada, et al \cite{wadaArxiv}.  

The density calculated is the local density felt by protons emitted from the PLS.  In a liquid-gas type phase transition, the protons represent the gas or low density region, which explains why the density is increasing with excitation energy.  At low excitation it is reasonable to expect that most fragments are ejected through a sequential surface type emission, i.e. at very low densities on the surface.  As we move to higher excitation, more nucleons get excited from the liquid to the gas leading to a higher local density felt by the protons.  Similarly, we would expect the density of the liquid to decrease with increasing event excitation.

%%%-----------------------------------------------------------------------------
%%% Fig. 3
%%%---------
\begin{figure}[tbp]
\includegraphics[width=0.50\textwidth, angle=0]{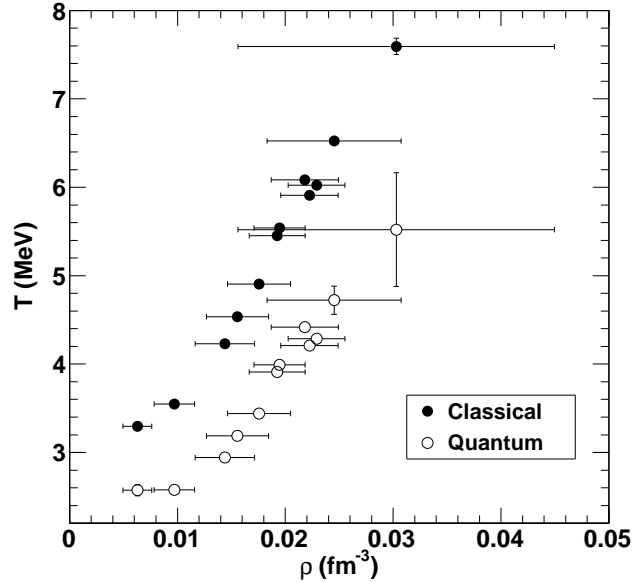}
\caption{Temperature versus density is plotted for both the classical and quantum temperature methods.  A direct relation can be seen with an increase in density corresponding to an increase in the difference in calculated temperature between the two methods.}
\label{fig:denstemp}
\end{figure}
%%%-------------------------------------------------------------------------------

This evolution from surface emission to volume emission may also explain features in both the normalized multiplicity and temperature plots.  The effect of the Pauli blocking responsible for the quenching observed in Fig. \ref{fig:mult} should be reduced with increasing event excitation since this will reduce the probability of two particles being at the same momentum.  However, increases in density increase the probability of interaction between particles; thus these two effects are competing processes.  This could explain why the normalized proton multiplicity increases only slightly from 2.5 to 7 MeV in event excitation.  It is possible that a similar argument could be made for the slow onset of quenching in the case of the A=3 particles.  With increasing event excitation, the probability of emitting composite particles increases, leading to a greater chance of such particles interacting, and therefore a greater effect due to Pauli blocking.  However, the extension to such cases is not straightforward and will be the subject of future investigations.  The effect of the density on the temperature is more easily seen by plotting these quantities against each other as shown in Fig. \ref{fig:denstemp}.  The increase in the difference between the two temperature calculations with increasing density is easily seen, and is attributed to the increase in Pauli blocking with increasing density.

%%%-----------------------------------------------------------------------------
%%% Fig. 4
%%%---------
\begin{figure}[tbp]
\includegraphics[width=0.50\textwidth, angle=0]{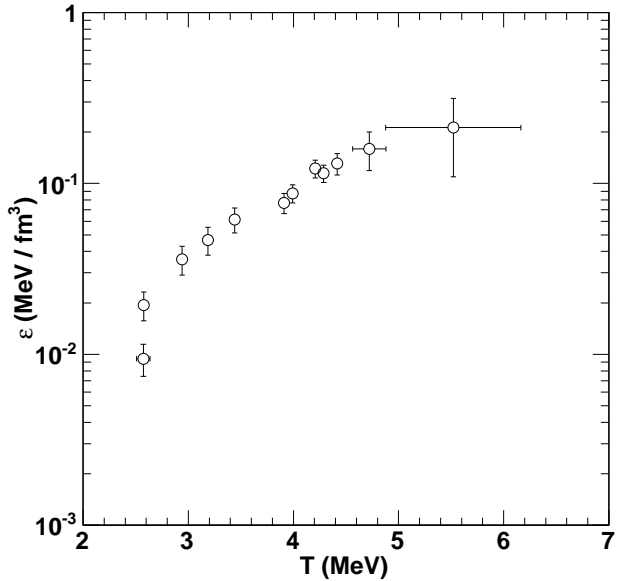}
\caption{Energy density versus temperature is shown.  The downturn in the data points around 2.5 MeV could indicate a region of rapid variation in the energy density indicative of a phase transition as predicted in Ref. \protect\cite{zheng2011}.  However, the unavailability of lower temperature points makes this inconclusive in this data set.  Statistical error is shown.}
\label{fig:edens}
\end{figure}
%%%-------------------------------------------------------------------------------

Variation in the density could also explain the lack of a noticeable plateau or flattening in the event excitation vs temperature often referred to as the caloric curve.  A flattening of the caloric curve is a feature of a system passing through a phase transition.  This feature may be obscured for systems whose density is not constant over a range of temperature.  For such as system, it may be instructive to compare the energy density versus the temperature.  Fig. \ref{fig:edens} shows the energy density $\varepsilon=<E_{xy}^{*}/A> \rho$ as a function of temperature.  A similar result was obtained in Ref. \cite{zheng2011}, suggesting a liquid-gas phase transition near 3 MeV due to a rapid variation of the energy density at that temperature.  Fig. \ref{fig:edens} agrees qualitatively with the data from Ref. \cite{zheng2011}; however, the rapid variation predicted falls at the lower limit of the experimental data.  We stress that the system discussed in Ref.[4] was much larger (A=80) compared to the one selected in this work (A=32) and therefore only a qualitative comparison is possible.

In conclusion, the first results of a new quantum method for the calculation of nuclear temperature and density have been presented.  The temperatures calculated show a substantial reduction when compared to a similarly derived classical method.  This reduction in temperature is in agreement with previous theoretical predictions showing that classically derived methods over predict nuclear temperature by failing to treat the Fermi motion of nucleons.  Quenching of the normalized multiplicity due to Pauli blocking has been observed for three species of light fermions.  Together these results highlight the need for a quantum treatment when dealing with statistical properties of fragmenting heavy ions.

%
% If in two-column mode, this environment will change to single-column
% format so that long equations can be displayed. Use
%\begin{widetext}
%\end{widetext}

% figures should be put into the text as floats.
% Use the graphics or graphicx packages (distributed with LaTeX2e)
% and the \includegraphics macro defined in those packages.
% See the LaTeX Graphics Companion by Michel Goosens, Sebastian Rahtz,
% entire page. There is no need to do explicit centering.

% \begin{figure}
% \end{figure}

%\appendix
%\section{}

\begin{acknowledgments}
The authors would like to thank the Cyclotron Institute staff for the excellent quality beam.  This work was supported in part by the Department of Energy through grant DE-FG03-93ER40773 and the Robert A. Welch Foundation through grant A-1266.
\end{acknowledgments}
\bibliography{TempPRL}   %%% the full filename is: TempPRL.bib

\begin{thebibliography}{17}%
\makeatletter
\providecommand \@ifxundefined [1]{%
 \@ifx{#1\undefined}
}%
\providecommand \@ifnum [1]{%
 \ifnum #1\expandafter \@firstoftwo
 \else \expandafter \@secondoftwo
 \fi
}%
\providecommand \@ifx [1]{%
 \ifx #1\expandafter \@firstoftwo
 \else \expandafter \@secondoftwo
 \fi
}%
\providecommand \natexlab [1]{#1}%
\providecommand \enquote  [1]{``#1''}%
\providecommand \bibnamefont  [1]{#1}%
\providecommand \bibfnamefont [1]{#1}%
\providecommand \citenamefont [1]{#1}%
\providecommand \href@noop [0]{\@secondoftwo}%
\providecommand \href [0]{\begingroup \@sanitize@url \@href}%
\providecommand \@href[1]{\@@startlink{#1}\@@href}%
\providecommand \@@href[1]{\endgroup#1\@@endlink}%
\providecommand \@sanitize@url [0]{\catcode `\\12\catcode `\$12\catcode
  `\&12\catcode `\#12\catcode `\^12\catcode `\_12\catcode `\%12\relax}%
\providecommand \@@startlink[1]{}%
\providecommand \@@endlink[0]{}%
\providecommand \url  [0]{\begingroup\@sanitize@url \@url }%
\providecommand \@url [1]{\endgroup\@href {#1}{\urlprefix }}%
\providecommand \urlprefix  [0]{URL }%
\providecommand \Eprint [0]{\href }%
\providecommand \doibase [0]{http://dx.doi.org/}%
\providecommand \selectlanguage [0]{\@gobble}%
\providecommand \bibinfo  [0]{\@secondoftwo}%
\providecommand \bibfield  [0]{\@secondoftwo}%
\providecommand \translation [1]{[#1]}%
\providecommand \BibitemOpen [0]{}%
\providecommand \bibitemStop [0]{}%
\providecommand \bibitemNoStop [0]{.\EOS\space}%
\providecommand \EOS [0]{\spacefactor3000\relax}%
\providecommand \BibitemShut  [1]{\csname bibitem#1\endcsname}%
\let\auto@bib@innerbib\@empty
%</preamble>
\bibitem [{\citenamefont {Bauer}(1995)}]{bauer1995}%
  \BibitemOpen
  \bibfield  {author} {\bibinfo {author} {\bibfnamefont {W.}~\bibnamefont
  {Bauer}},\ }\href@noop {} {\bibfield  {journal} {\bibinfo  {journal} {Phys.
  Rev. C}\ }\textbf {\bibinfo {volume} {51}},\ \bibinfo {pages} {803} (\bibinfo
  {year} {1995})}\BibitemShut {NoStop}%
\bibitem [{\citenamefont {Morrissey}\ \emph {et~al.}(1994)\citenamefont
  {Morrissey}, \citenamefont {Benenson},\ and\ \citenamefont
  {Friedman}}]{morrissey1994}%
  \BibitemOpen
  \bibfield  {author} {\bibinfo {author} {\bibfnamefont {D.~J.}\ \bibnamefont
  {Morrissey}}, \bibinfo {author} {\bibfnamefont {W.}~\bibnamefont {Benenson}},
  \ and\ \bibinfo {author} {\bibfnamefont {W.~A.}\ \bibnamefont {Friedman}},\
  }\href@noop {} {\bibfield  {journal} {\bibinfo  {journal} {Annu. Rev. Nucl.
  Part. Sci}\ }\textbf {\bibinfo {volume} {44}},\ \bibinfo {pages} {27}
  (\bibinfo {year} {1994})}\BibitemShut {NoStop}%
\bibitem [{\citenamefont {Keli\'{c}}\ \emph {et~al.}(2006)\citenamefont
  {Keli\'{c}}, \citenamefont {Natowitz},\ and\ \citenamefont
  {Schmidt}}]{kelic2006}%
  \BibitemOpen
  \bibfield  {author} {\bibinfo {author} {\bibfnamefont {A.}~\bibnamefont
  {Keli\'{c}}}, \bibinfo {author} {\bibfnamefont {J.}~\bibnamefont {Natowitz}},
  \ and\ \bibinfo {author} {\bibfnamefont {K.-H.}\ \bibnamefont {Schmidt}},\
  }\href@noop {} {\bibfield  {journal} {\bibinfo  {journal} {Eur. Phys. J. A}\
  }\textbf {\bibinfo {volume} {30}},\ \bibinfo {pages} {203} (\bibinfo {year}
  {2006})}\BibitemShut {NoStop}%
\bibitem [{\citenamefont {Zheng}\ and\ \citenamefont
  {Bonasera}(2011{\natexlab{a}})}]{zheng2011}%
  \BibitemOpen
  \bibfield  {author} {\bibinfo {author} {\bibfnamefont {H.}~\bibnamefont
  {Zheng}}\ and\ \bibinfo {author} {\bibfnamefont {A.}~\bibnamefont
  {Bonasera}},\ }\href@noop {} {\bibfield  {journal} {\bibinfo  {journal}
  {Phys. Lett. B}\ }\textbf {\bibinfo {volume} {696}},\ \bibinfo {pages} {178}
  (\bibinfo {year} {2011}{\natexlab{a}})}\BibitemShut {NoStop}%
\bibitem [{\citenamefont {Zheng}\ and\ \citenamefont
  {Bonasera}(2011{\natexlab{b}})}]{zhengArxiv}%
  \BibitemOpen
  \bibfield  {author} {\bibinfo {author} {\bibfnamefont {H.}~\bibnamefont
  {Zheng}}\ and\ \bibinfo {author} {\bibfnamefont {A.}~\bibnamefont
  {Bonasera}},\ }\href@noop {} {} (\bibinfo {year} {2011}{\natexlab{b}}),\
  \Eprint {http://arxiv.org/abs/nucl-th/1105.0563} {nucl-th/1105.0563}
  \BibitemShut {NoStop}%
\bibitem [{\citenamefont {Muller}\ \emph {et~al.}(2010)\citenamefont {Muller}
  \emph {et~al.}}]{muller2010}%
  \BibitemOpen
  \bibfield  {author} {\bibinfo {author} {\bibfnamefont {T.}~\bibnamefont
  {Muller}} \emph {et~al.},\ }\href@noop {} {\bibfield  {journal} {\bibinfo
  {journal} {Phys. Rev. Lett.}\ }\textbf {\bibinfo {volume} {105}},\ \bibinfo
  {pages} {040401} (\bibinfo {year} {2010})}\BibitemShut {NoStop}%
\bibitem [{\citenamefont {Sanner}\ \emph {et~al.}(2010)\citenamefont {Sanner}
  \emph {et~al.}}]{sanner2010}%
  \BibitemOpen
  \bibfield  {author} {\bibinfo {author} {\bibfnamefont {C.}~\bibnamefont
  {Sanner}} \emph {et~al.},\ }\href@noop {} {\bibfield  {journal} {\bibinfo
  {journal} {Phys. Rev. Lett.}\ }\textbf {\bibinfo {volume} {105}},\ \bibinfo
  {pages} {040402} (\bibinfo {year} {2010})}\BibitemShut {NoStop}%
\bibitem [{\citenamefont {Westbrook}(2010)}]{westbrook2010}%
  \BibitemOpen
  \bibfield  {author} {\bibinfo {author} {\bibfnamefont {C.}~\bibnamefont
  {Westbrook}},\ }\href@noop {} {\bibfield  {journal} {\bibinfo  {journal}
  {Physics}\ }\textbf {\bibinfo {volume} {3}},\ \bibinfo {pages} {39} (\bibinfo
  {year} {2010})}\BibitemShut {NoStop}%
\bibitem [{\citenamefont {Bonasera}(2000)}]{bonasera2000}%
  \BibitemOpen
  \bibfield  {author} {\bibinfo {author} {\bibfnamefont {A.}~\bibnamefont
  {Bonasera}},\ }\href@noop {} {\bibfield  {journal} {\bibinfo  {journal}
  {Phys. Rev. C}\ }\textbf {\bibinfo {volume} {62}},\ \bibinfo {pages}
  {052202(R)} (\bibinfo {year} {2000})}\BibitemShut {NoStop}%
\bibitem [{\citenamefont {Papa}\ \emph {et~al.}(2001)\citenamefont {Papa} \emph
  {et~al.}}]{papa2001}%
  \BibitemOpen
  \bibfield  {author} {\bibinfo {author} {\bibfnamefont {M.}~\bibnamefont
  {Papa}} \emph {et~al.},\ }\href@noop {} {\bibfield  {journal} {\bibinfo
  {journal} {Phys. Rev. C}\ }\textbf {\bibinfo {volume} {64}},\ \bibinfo
  {pages} {024612} (\bibinfo {year} {2001})}\BibitemShut {NoStop}%
\bibitem [{\citenamefont {Gimeno-Nogues}\ \emph {et~al.}(1997)\citenamefont
  {Gimeno-Nogues} \emph {et~al.}}]{gimeno97}%
  \BibitemOpen
  \bibfield  {author} {\bibinfo {author} {\bibfnamefont {F.}~\bibnamefont
  {Gimeno-Nogues}} \emph {et~al.},\ }\href@noop {} {\bibfield  {journal}
  {\bibinfo  {journal} {Nucl. Inst. and Meth. A}\ }\textbf {\bibinfo {volume}
  {404}},\ \bibinfo {pages} {470} (\bibinfo {year} {1997})}\BibitemShut
  {NoStop}%
\bibitem [{\citenamefont {Stein}\ \emph {et~al.}(2009)\citenamefont {Stein}
  \emph {et~al.}}]{stein2009}%
  \BibitemOpen
  \bibfield  {author} {\bibinfo {author} {\bibfnamefont {B.~C.}\ \bibnamefont
  {Stein}} \emph {et~al.},\ }\href@noop {} {\bibfield  {journal} {\bibinfo
  {journal} {AIP Conf. Proc.}\ }\textbf {\bibinfo {volume} {1099}},\ \bibinfo
  {pages} {700} (\bibinfo {year} {2009})}\BibitemShut {NoStop}%
\bibitem [{\citenamefont {Keksis}\ \emph {et~al.}(2010)\citenamefont {Keksis}
  \emph {et~al.}}]{keksis2010}%
  \BibitemOpen
  \bibfield  {author} {\bibinfo {author} {\bibfnamefont {A.~L.}\ \bibnamefont
  {Keksis}} \emph {et~al.},\ }\href@noop {} {\bibfield  {journal} {\bibinfo
  {journal} {Phys. Rev. C}\ }\textbf {\bibinfo {volume} {81}} (\bibinfo {year}
  {2010})}\BibitemShut {NoStop}%
\bibitem [{\citenamefont {Soisson}\ \emph {et~al.}(2011)\citenamefont {Soisson}
  \emph {et~al.}}]{soisson2011}%
  \BibitemOpen
  \bibfield  {author} {\bibinfo {author} {\bibfnamefont {S.~N.}\ \bibnamefont
  {Soisson}} \emph {et~al.},\ }\href@noop {} {\bibfield  {journal} {\bibinfo
  {journal} {Nucl. Phys. A}\ } (\bibinfo {year} {2011})},\ \bibinfo {note}
  {(Submitted)}\BibitemShut {NoStop}%
\bibitem [{\citenamefont {Wuenschel}\ \emph {et~al.}(2010)\citenamefont
  {Wuenschel} \emph {et~al.}}]{wuenschel2010}%
  \BibitemOpen
  \bibfield  {author} {\bibinfo {author} {\bibfnamefont {S.}~\bibnamefont
  {Wuenschel}} \emph {et~al.},\ }\href@noop {} {\bibfield  {journal} {\bibinfo
  {journal} {Nucl. Phys. A}\ }\textbf {\bibinfo {volume} {843}},\ \bibinfo
  {pages} {1} (\bibinfo {year} {2010})}\BibitemShut {NoStop}%
\bibitem [{\citenamefont {Wuenschel}(2009)}]{wuenschelThesis}%
  \BibitemOpen
  \bibfield  {author} {\bibinfo {author} {\bibfnamefont {S.}~\bibnamefont
  {Wuenschel}},\ }\href@noop {} {Ph.D. thesis},\ \bibinfo  {school} {Texas A\&M
  University} (\bibinfo {year} {2009})\BibitemShut {NoStop}%
\bibitem [{\citenamefont {Wada}\ \emph {et~al.}(2011)\citenamefont {Wada} \emph
  {et~al.}}]{wadaArxiv}%
  \BibitemOpen
  \bibfield  {author} {\bibinfo {author} {\bibfnamefont {R.}~\bibnamefont
  {Wada}} \emph {et~al.},\ }\href@noop {} {} (\bibinfo {year} {2011}),\ \Eprint
  {http://arxiv.org/abs/nucl-ex/1110.3341} {nucl-ex/1110.3341} \BibitemShut
  {NoStop}%
\end{thebibliography}%
\end{document}